\documentclass[doublecol]{epl2}

\title{Bilinear Effect in Complex Systems}
\shorttitle{Title} 

\author{Lui Lam\inst{1} \and David C. Bellavia\inst{1} \and Xiao-Pu Han\inst{2} \and Chih-Hui Alston Liu\inst{1} \and Chang-Qing Shu\inst{3} \and Zhengjin Wei\inst{4} \and Tao Zhou\inst{2,5} \and Jichen Zhu\inst{6}}
\shortauthor{L. Lam \etal}

\institute{
  \inst{1} Department of Physics and Astronomy, San Jose State University, San Jose, CA 95192-0106, U.S.A.\\
  \inst{2} Department of Modern Physics, University of Science and Technology of China, Hefei 230026, China\\
  \inst{3} ADACEL Systems Incorporation, 5945 Hazeltine National Drive, Orlando, FL 32822, U.S.A. \\
  \inst{4} Nanjing Municipal Museum, 4 Chao Tian Gong, Nanjing 210004, China \\
  \inst{5} Web Sciences Center, University of Electronic Science and Technology of China, Chengdu 610054, China\\
  \inst{6} School of Literature, Communication, and Culture, Georgia Institute of Technology, 686 Cherry St., Atlanta, GA 30332-0165, U.S.A.}

\abstract{
The distribution of the lifetime of Chinese dynasties (as well as that of the British Isles and Japan) in a linear Zipf plot is found to consist of two straight lines intersecting at a transition point. This two-section piecewise-linear distribution is different from the power law or the stretched exponent distribution, and is called the \textit{Bilinear Effect} for short. With assumptions mimicking the organization of ancient Chinese regimes, a 3-layer network model is constructed. Numerical results of this model show the bilinear effect, providing a plausible explanation of the historical data. Bilinear effect in two other social systems is presented, indicating that such a piecewise-linear effect is widespread in social systems.
}

\begin{document}

\maketitle

\section{Introduction}

A common way to characterize and classify complex systems is through the Zipf plots \cite{Zipf}.  Given a sequence of numbers, the corresponding Zipf plot is obtained in four steps: (i) The sequence is rearranged in a decreasing order. (ii) For numbers of the same magnitude, retain only one of them in the sequence. (iii) The largest number is assigned rank 1, the second largest rank 2, etc. (iv) The Zipf plot is the curve of number vs. rank (R). Note that as a result of the decreasing order, the Zipf plot so defined is always a monotonically decreasing curve. Since the Zipf plot can give an inverse function of the cumulative distribution from original data \cite{New}, it is widely used in the statistical analysis of small samples \cite{Reed, Hxp}.

There are two well-known non-Poisson types of Zipf plots: power laws \cite{Zipf, New} and stretched exponents \cite{Lah}. The Power law distribution has been widely observed in a large number of self-organizing systems. In the last several decades the power law distribution has attracted the attention of many scientists. It is at the center of complex systems research because of its special mathematical and dynamical properties\cite{Alb,Newm,Pas}, and physical implications\cite{bak}. In the last decade, the rise of research in the network sciences makes the power law more prominent \cite{Wat1,Bar1}.

The stretched exponent distribution generally can be viewed as an intermediate form between a scaling form (power law) and homogenous types of distributions (such as Poisson distribution). It has also been widely observed in many social and material systems \cite{XB,Hol,Str,Han}. The typical forms of both the power law and stretched exponent is that of a continuous curve in a Zipf plot.

Here we investigate the rank distributions of the lifetime of the dynasties in ancient China, British Isles and Japan. The surprising result is that these distributions in the linear Zipf plots obey neither power-law nor stretched exponent type, but a special two-section piecewise-linear function. The two monotonic decreasing straight lines intersect at a transition point; the slope of the curve is not continuous. In the rest of this paper, this type of two-section piecewise-linear form in a linear Zipf plot is called Bilinear Effect for short. Previously, multiple piecewise-linear forms have been discussed as a phenomena or an assumption in many natural and technological systems \cite{Sin,Bat,Chan,Carb,Thom}. For example, Ref. \cite{Sin} and \cite{Bat} have respectively studied the stochastic resonance and synchronization dynamics in piecewise linear maps; Ref. \cite{Chan} has analyzed the economic problem in the distribution of products in supply chain with piecewise-linear cost structures;  Ref. \cite{Carb} has investigated the discontinuity correction in piecewise-linear models of oscillators; and Ref. \cite{Thom} has proposed a random number generator based on piecewise-linear approximations. In this regard, bilinear effect is a special piecewise-linear form ----- a two-section version; our findings indicates that it widely exists in complex systems, specially in social systems. Below, we present a dynamical model to explain its possible underlying mechanism, and introduce two more examples of such bilinear rank distribution in other social systems.



\section{Bilinear effect in the lifetime of dynasties}

\begin{figure}
  \includegraphics[width=8.8cm]{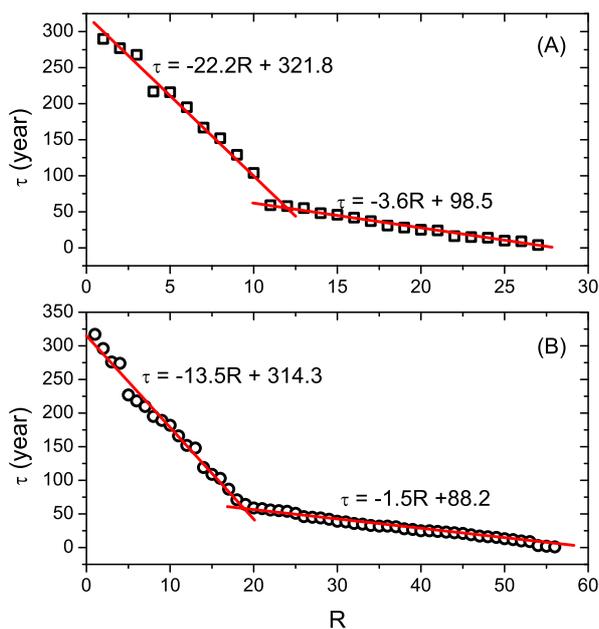}\\
  \caption{(Color online) Lifetime distribution of Chinese dynasties in linear Zipf plot. (A) data from Ref. \cite{Mor}, the corresponding dynasty of each data point is (from left to right):  Tang,
Ming, Qing, Liao, Western Han, Eastern Han, Northern
Sung, Northern Wei (abreast of Southern Sung and Yuan),
Chin, Eastern Chin, Liu Sung, Wu, Liang, Western Chin, Wei,
Minor Han (abreast of Western Wei), Sui, Chen, Northern
Chi, Northern Zhou, Southern Chi, Hsin (abreast of Later
Liang), Qin, Later Tang, Later Chin, Later Zhou, Later
Han.  (B) data from Ref. \cite{Cihai}.}
\end{figure}

\begin{figure}
  \includegraphics[width=8.8cm]{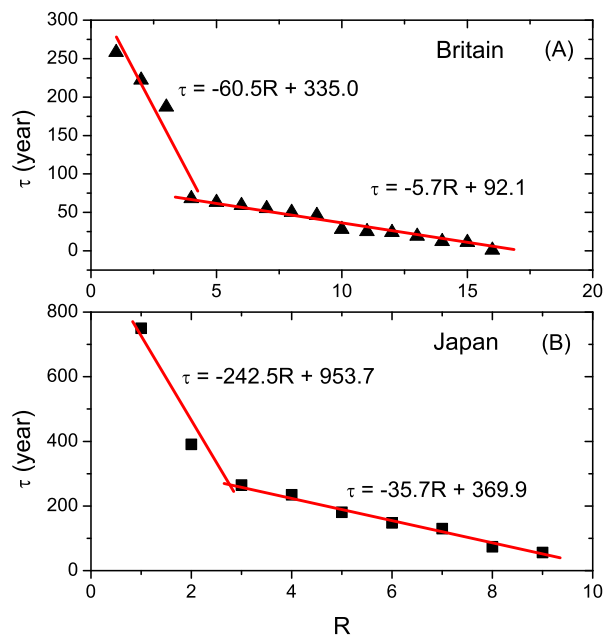}\\
  \caption{(Color online) Lifetime distributions of dynasties of Britain (A)  and Japan (B). Data from Ref. \cite{Mor}.}
\end{figure}


The dynasty cycle in Chinese history has been studied mainly by nonlinear dynamical models in the last two decades. One of the features noticed by these early researches is the periodic alternation of society between despotism and anarchy. Usher proposed a differential model based on the interaction between three basic classes in ancient China: farmers, bandits and rulers \cite{Ush}. This model includes three simple three-variable differential equations, denoting, respectively, the evolution of the three classes. This model successfully generated the alternation between despotism and anarchy, which is similar to the basic feature in dynasties cycle. Based on this work, several differential game models were proposed \cite{Fei1, Chu, Fei}, and the generalized version of this class of models was further investigated \cite{Gro}; some of them extended the discussions to the evolution of population in ancient China \cite{Chu}. Most of these works paid much attention to the nonlinear dynamical properties of these models, but the statistical pattern of real-world dynasty cycles are rarely discussed. Here in this paper, the distribution of lifetime of Chinese dynasties is investigated for the first time, to the best of our knowledge. 

The record of the history of each dynasty in ancient China from Qin Dynasty (221 B.C.) to Qing Dynasty is exhaustive and reliable. The Zipf plots of the lifetimes of these dynasties are presented in Fig. 1 (A) and (B).
Both of the two sets of data range from Qin to Qing dynasty (221 B.C. to 1912
A.D.).  The data for Fig. 1 (A) is obtained from Ref. \cite{Mor}, which includes
31 main Chinese dynasties. The data for Fig. 1 (B) is from Cihai \cite{Cihai}, a Chinese encyclopedia;  which not only includes the 31 main dynasties, but also many local powers and provisional governments, resulting in a total of 74 dynasties. The two data sets are listed on the website \cite{List}. These two sets of data depict similar behavior -----  the decreasing two-section piecewise-linear form, or, the bilinear effect. The transition point in these two Zipf plots is  $\tau = 57 \pm 2$ years. This implies that if a Chinese dynasty survives longer than 57 years, it will have a greater chance of
surviving longer, and the chance that it will be destroyed is sharply reduced. For example, Fig. 1(A) implies that a dynasty
can survive $3.6\pm0.1$ years if it lasts $57\pm2$ years or less; beyond that, every $22.2 \pm 0.1$ years. In other words, the distribution of the lifetimes of Chinese dynasties is discrete, or ``quantized". Moreover, this is the phenomenon that a human
entity, a dynasty in this case, becomes stronger or more stable after existing for a period of time. The mere
fact of survival reinforces its strength, through adaptive learning, restructuring, or other means.

Bilinear effect can also be observed in the lifetime distribution of dynasties of some other countries. Fig. 2 shows two examples: the dynasty lifetime distributions of Britain and Japan. In contrast to the Chinese data, the transition point of British and Japan, respectively, is $\tau = 68 \pm 2$ and $268 \pm 10$, which are larger than that of China. However, the number of data points for these countries are less than those of China, and this is why the bilinear effect is less certain in these
systems.

In the following discussion, to understand the underlying mechanism, a governmental structure giving rise to the bilinear effect is introduced.

\section{The 3-layer network model}

Roughly speaking, the government structure of a Chinese regime in the last two thousand or so years since the Qin dynasty consists of three layers: the emperor court (the central government), the provinces, and the cities/villages. They are represented schematically by layer $A$, $B$ and $C$, respectively, in Fig. 3 here.

Every year, the cities/villages submit part of their income, in the form of ``taxes" to the upper layer, the provincial governments. And the provincial governments in turn submit a certain amount of their revenues to the emperor court, the top layer. At the same time, the emperor court maintains its control by allocating funds/resources to the governments in the middle layer as it pleases. But there will be no downward flow of resources from layer $B$ to layer $C$. The communication between local governments in each layer is indirect and will be ignored in our model.

\begin{figure}
  \includegraphics[width=8.8cm]{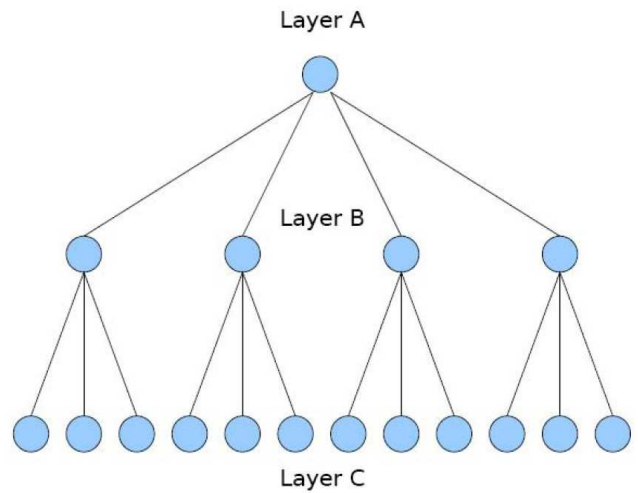}\\
  \caption{(Color online) The structure of the 3-layer network model ($N_B = 4$ and $N_C = 3$).}
\end{figure}

In our 3-layer network model, the upper layer A has one node. Node $A$ is connected to all the nodes $B_{i}$ $(i = 1, \cdots , N_B)$ in the second layer (layer $B$). And each node $B_i$ is connected to nodes $C_{ij }$ $(j = 1, \cdots , N_C)$ in the third layer (layer $C$). So the number of nodes on layer $B$ and layer $C$, respectively, is $N_B$ and $N_C$. There is no connection between the nodes in the same layer.

Let $F(B_i, t)$ be the amount of resources (or fitness) possessed by node $B_i$ at time $t$; similarly $F(A, t)$ for node $A$, and $F(C_{ij}, t)$ for node $C_{ij}$. Each node will transfer part of its own resource to the other nodes that are connected to it, according to the following rules.

(i) From $A$ to $B_i$: At time $t$, a node in layer $B$ ($k$, say) is random picked and an amount $T_A(t)$ is transferred from node $A$ to node $B_k$ such that
\begin{equation}
	T_A(t) = aF(B_k, t - 1)
\end{equation}

(ii) From $B_i$ to $A$: An amount $T_{BA}$ is transferred from node $B_i$ to node $A$ such that
\begin{equation}
	T_{BA}(B_i, A, t) = bF(B_i, t - 1);
\end{equation}

(iii) From $C_{ij}$ to $B_i$: An amount $T_{CB}$ is transferred from node $C_{ij}$ to node $B_i$ such that
\begin{equation}
	T_{CB}(C_{ij}, B_i, t) = cF(C_{ij}, t - 1).
\end{equation}
where $a$, $b$ and $c$ are constants (each one is less than 1).

To keep itself running, node $A$ does consume its own resources; the amount is denoted by $eF(A, t)$ with $e (< 1)$ being a constant. It follows that the time evolution of the fitness at each node is given by

(i) For node A:
\begin{equation}
\begin{array}{l}
F(A, t) - F(A, t - 1) \\
= - T_A(t) + \sum_iT_{BA}(B_i, A, t) - eF(A, t - 1)
\end{array}
\end{equation}

(ii) For node $B_i$:
\begin{equation}
\begin{array}{l}
F(B_i, t) - F(B_i, t - 1)\\ = T_A(t)\delta_{ik} + \sum_j T_{CB}(C_{ij}, B_i, t) - T_{BA}(B_i, A, t)
\end{array}
\end{equation}

(iii) For node $C_{ij}$:
\begin{equation}
	F(C_{ij}, t) - F(C_{ij}, t - 1) = - T_{CB}(C_{ij}, B_i, t).
\end{equation}

Starting with initial fitness for the nodes, the computer run is stopped when $F(A, t) \leq 0$ for the first time (at time $t = \tau$, say), which mimics the exhaustion of the resources of the central government. The lifetime of the regime is taken to be $\tau$.

\begin{figure}
  \includegraphics[width=8.8cm]{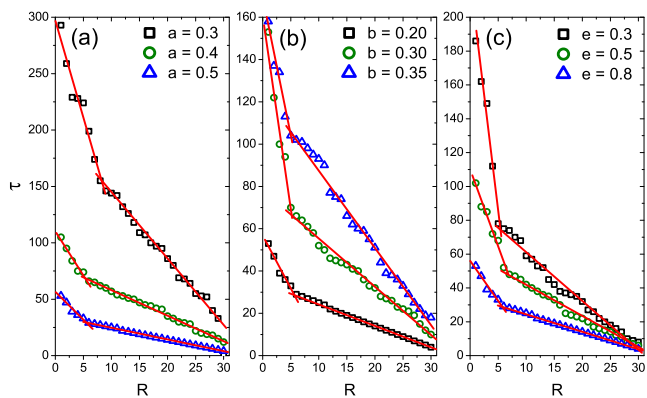}\\
  \caption{ (Color online) Zipf plot resulted from a particular pick ($N_p = 30$) generated by the 3-layer network model ($N_B = 4$, $N_C = 3$, and $c = 0.3$). (a) Results for different $a$; other parameters used: $b = 0.2$, and $e = 0.8$. (b) Results for different $b$; other parameters used: $a = 0.5$, and $e = 0.8$. (c) Results for different $e$; other parameters used: $a = 0.5$, and $b = 0.2$. The initial conditions are: $F(A, 0) = 100$, $F(B_i, 0) = 50$ and $F(C_{ij}, 0) = 30$. These results show bilinear effect
similar to that in Fig. 1(A). }
\end{figure}

For a set of given parameters, many computational runs of this model are performed, giving rise to the normalized probability function $p(\tau)$ ------ such that  $p(\tau)$ is the probability that $\tau$ is found among all the runs. A sequence of numbers, $\lbrace \tau_i \rbrace$ with $i = 1, 2, \cdots , N_p$, are picked according to this $p(\tau)$.  The Zipf plots derived from particular sequences so picked are depicted in Fig. 4. All of these obviously show bilinear effect, indicating that this property can indeed emerge from the government resource assignment process. We further investigate the impacts of each parameter on the bilinear effect. The main parameters of our model include $N_B$, $N_C$, $a$, $b$, $c$ and $e$. Simulation results indicate that this model can generate bilinear $\tau$ rank distribution in a wide parameter settings. What we focus on here is the value of the transition point.


The value of $\tau$ of the transition point are sensitively impacted by parameter $a$, $b$ and $e$. As show in Fig. 4, $\tau$ of the transition point increases along with the reduction of $a$ and $e$, and the rise of $b$.  Parameter $c$ has no obvious impact on the transition point. Parameter $a$ and $b$ denote the strength of the transport of resources between the central government and several local powers, and $e$ denotes the consumption of the central government; and thus the transition point is mainly impacted on by the resource flow in the upper layers. This conclusion could be a key to understand the difference of the transition point in different countries. In parallel, when $a>0.5$, $b<0.2$  or $e>0.8$, the bilinear effect is difficult to emerge in our model. In contrast, the impact of other parameters ($N_B$, $N_C$) on the transition point is insensitive. 

\section{Other examples of bilinear effect}

Except for the distribution of lifetime of dynasties, two other interesting examples of such bilinear effect in
social systems are found.


One example is the number of online votes for Chinese \emph{Xiaopin} actors (\emph{Xiaopin} is a popular form of short drama performed by a cast of usually two actors in China; the data is available from http://ent.sin.com.cn/2004-09-30/1050521359.html (Oct. 7, 2004)). In this vote, each voter can choose their favorite actor from a list of 30. The more number of votes an actor gets, the higher the popularity. The bilinear effect shows obviously in the plot in Fig. 5 (A). This implies these actors can be divided into two groups. In other words, the social reputation of actors could be dichotomous: if an actor can pass a critical popularity, he/she will achieve greater popularity more easily. This result contradicts the common understanding that the social reputation of people is continuously distributed.

Another example is the 2004 airline quality ratings (data available from http://www.aqr.aero/aqrreports/2005aqr.pdf), as shown in Fig. 5 (B). These examples imply that that bilinear effect could be widespread in some complex systems, especially in social systems.

\begin{figure}
  \includegraphics[width=8.8cm]{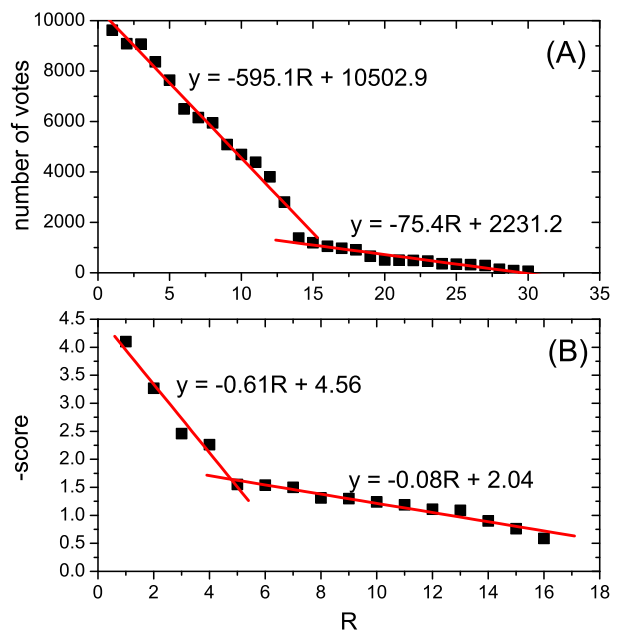}\\
  \caption{(Color online) Two examples of bilinear effect. A. Online popular votes for \emph{xiaopin} actors. B. Airline quality data. In both plots "y" in the equation corresponds to the quantity on the vertical axis.}
\end{figure}

\section{Conclusions}

We present examples from three different types of social systems that show
a special piecewise-linear distribution called the bilinear effect.

The most interesting characteristic
of bilinear effect is that the samples are divided into two distinct linearly distributed groups which are connected by
a sharp transition point. It indicates that the statistical
properties of some complex systems could be discrete.
What is the meaning of the transition point? It could be an indication of a ``phase transition".
However, this supposition needs more empirical evidence and theoretical
understanding.

Our research of the bilinear effect covers several different
realms, including the lifetimes of dynasties of several countries, online votes of actors, and airline quality ratings.
These empirical results imply that bilinear effect could has some universal significance.

We propose a 3-layer network model to investigate the
underlying mechanisms of the bilinear effect in the lifetimes
of dynasties. This model can generate bilinear effect
in wide parameter settings, in agreement with our
empirical data. This model also makes it possible to understand the difference
of the transition point in different countries. While this model does provide a plausible explanation of the origin of the bilinear effect in dynasty lifetimes, it is not a microscopic model, and it does not explain the bilinear effect in other social systems.. A more sophisticated model is needed in all these cases. Other models and explanations are not precluded a priority. The fact that there could be more than one
mechanism in producing the bilinear effect is not that surprising.
A similar case exists in the case of power
laws in Zipf plots \cite{New}.

In summary, we report the empirical statistics and modeling
for the bilinear effect in Zipf plots in several social
systems. The dynasty lifetime results contribute to the advancement of histophysics \cite{Lam}.  Although this letter investigates a few examples, the research is just at its beginning. There are still many open questions that will require insightful research to understand.







\acknowledgments

XPH and TZ acknowledge the support of 973 program (2006CB705500),
and the National Natural Science Foundation of China (10532060, 10635040 and 70871082).



\begin{thebibliography}{0}

\bibitem{Zipf}	\Name{Zipf G. K.}, Human Behavior and the Principle of Least Effort (Addison-Wesley, Cambridge, MA, 1949).
\bibitem{New}	\Name{Newman M. E. J.} \REVIEW{Contemp. Phys.} {46}{2005}{323}.

\bibitem{Reed} \Name{Reed W.J.} \REVIEW{Economics Lett.} {74}{2001}{15}.
\bibitem{Hxp} \Name{Han X.-P., Wang B.-H., Zhou C.-S, Zhou T., and Zhu J.-F.}{arxiv: 0912.1390}.

\bibitem{Lah} \Name{Laherr\`{e}re J.  and Sornette D.} \REVIEW{Eur. Phys. J. B}{2}{1998}{525}.

\bibitem{Alb} \Name{Albert R., Jeong H., and Barab\'{a}si A. -L.} \REVIEW{Nature}{406}{2000}{378}.
\bibitem{Newm} \Name{Newman M.E.J.},  \REVIEW{Phys Rev E}{66}{2002}{016128}.

\bibitem{Pas} \Name{Pastor-Satorras R., Vespignani A.}\REVIEW{Phys Rev Lett}{86}{2001}{3200}.
\bibitem{bak} \Name{Bak P., Tang C., and Wiesenfeld K.} \REVIEW{Phys. Rev. Lett}{59} {1987} {381}.
\bibitem{Wat1} \Name{Watts D. J., Strogatz S. H.} \REVIEW{Nature}{393}{1998}{440}.
\bibitem{Bar1} \Name{Barab\'{a}si A. -L., Albert R.}\REVIEW{Science}{286}{1999}{509}.
\bibitem{XB} \Name{Xulvi-Brunet R., and Sokolov I. M.} \REVIEW{Phys. Rev. E}{66}{2002}{026118}.
\bibitem{Hol} \Name{Holanda A. J., Pisa I. T., Kinouchi O., Martinez A. S., and Ruiz E. E. S.} \REVIEW{Physica A}{344}{2004}{530}.
\bibitem{Str} \Name{Sturman B., Podivilov E., and Gorkunov M.} \REVIEW{Phys. Rev. Lett.}{91}{2005}{176602}.
\bibitem{Han} \Name{Han X.-P., Hu C.-D., Liu Z.-M., and Wang B.-H.}\REVIEW{Europhys. Lett.}{83}{2008}{28003}.


\bibitem{Sin} \Name{Sinha S., and Chakrabarti. B. K.} \REVIEW{Phys. Rev. E}{58}{1998}{8009}.
\bibitem{Bat} \Name{Batista A. M., Pinto S. E. de S., Viana R. L., and Lopes S. R.}  \REVIEW{Phys. Rev. E}{65}{2002}{056209}.
\bibitem{Chan} \Name{Chan L. M. A., Muriel A., Shen Z. -J. M., Simchi-Levi D., and Teo C. -P.}\REVIEW{Management Sci.}{48}{2002}{1446}.
\bibitem{Carb} \Name{Carbone A., and Palma F.}\REVIEW{Int. J. Circ. Theor. Appl.}{35}{2007}{93}.
\bibitem{Thom} \Name{Thomas D.B., and Luk W.}REVIEW{IET Comput. Digit. Tech.}{1}{2007}{312}.

\bibitem{Ush} \Name{Usher D.} \REVIEW{Am. Econ. Rev.}{79}{1989}{1031}.
\bibitem{Fei1} \Name{Feichtinger G., and Novak A. J.} \REVIEW{J. Optimiz. Theory and Appl.}{80}{1994}{407}.
\bibitem{Chu}  \Name{Chu C. Y. C., and Lee R. D. }\REVIEW{J. Popul. Econ}{7} {1994}{351}.
\bibitem{Fei}  \Name{Feichtinger G., Forst C. V., and Piccardi C.}\REVIEW{Chaos, Solitons and Fractals}{7} {1996}{257}.
\bibitem{Gro}  \Name{Gross T., Feudel U.}\REVIEW{Phys. Rev. E}{73}{2006}{016205}.



\bibitem{Mor}	\Name{Morby J. E.}  Dynasties of the World (Oxford U. P., Oxford, 2002).
\bibitem{Cihai} \Name{Xia Z.-N.} Cihai (Shanghai Lexicographical P. H., Shanghai, 1979).

\bibitem{List} {http://www.sciencenet.cn/upload/blog/file/2010/8/ 2010817181447879502.pdf}


\bibitem{Lam}  \Name{Lam L.}\REVIEW{Mod. Phys. Lett. B}{16}{2002}{1163}.







\end{thebibliography}
\end{document}